\documentstyle[twocolumn,prl,aps,graphicx]{revtex}

\begin{document}
\DeclareGraphicsExtensions{.eps,.jpg,.pdf,.mps,.png} \draft
\twocolumn[\hsize\textwidth\columnwidth\hsize\csname
@twocolumnfalse\endcsname
\title{Electric field induced memory and aging effects in pure solid N$_2$}
\author{S.\ Pilla, J.\ A.\ Hamida, K.\ A.\ Muttalib, and N.\ S.\ Sullivan}
\address{Department of Physics, University of Florida, Gainesville, FL 32611}
\date{\today}
\maketitle

\begin{abstract}
We report combined high sensitivity dielectric constant and heat
capacity measurements of pure solid N$_2$ in the presence of a
small external ac electric field in the audio frequency range. We
have observed strong field induced aging and memory effects which
show that field cooled samples may be prepared in a variety of
metastable states leading to a free energy landscape with
experimentally ``tunable'' barriers, and tunneling between these
states may occur within laboratory time scales.
\end{abstract}

] \narrowtext

The interplay of geometric frustration and strong ferromagnetic
as well as antiferromagnetic interactions in magnetic systems
such as Gd$_3$Ga$_5$O$_{12}$, Y$_2$Mo$_2$O$_7$, Ho$_2$Ti$_2$O$_7$
and SrCr$_{9p}$Ga$_{12-9p}$O$_{19}$ with $p\approx0.9$ have been
the subject of intense interest
\cite{Tsui,Petrenko,Gingras,Harris,Ramirez}. In these systems,
frustration is caused by competition between nearest neighbor
exchange interaction which leads to macroscopic degeneracies and
can lead to a variety of new phenomena at low temperatures
including spin glasses, spin liquids and magnetic analogs of ice.
However, a good understanding of these systems is hindered by the
fact that these systems are highly complicated, and the
interactions are often complex. In the present work we show that
there exists a simple electrical analog of such systems, namely
pure solid N$_2$, where the $hcp$ lattice geometry at
temperatures above $T_{\alpha\beta}$ is incompatible with the
symmetry of the interactions and the resulting geometrical
frustration destroys the long range orientational order favored
by the electric quadrupolar interactions between molecules. There
are several unique features that make solid N$_2$ an ideal system
to study the interplay of geometric frustration and interactions:
(i) It is a simple system with well understood molecular
interactions \cite{Scott,Manzhelii}. (ii) Solid N$_2$ undergoes
structural transition from $hcp$ to $fcc$ at $T_{\alpha\beta}$
which significantly lowers the geometric frustration; this gives
us the unique opportunity to study the effects of geometric
frustration by comparing the system in the two geometries. (iii)
As we will show in the present work, but unsuspected previously,
the system can be manipulated to become trapped in a variety of
metastable states by cooling it in a small ac electric field in
the audio frequency range. (iv) The time scale involved for the
system to tunnel between these trapped macroscopic states can be
several hours (short compared to silicate glasses), which allows
for the possibility of studying the nonequilibrium behavior
including aging effects in the laboratory\cite{Alberici,Franz}.
(v) It is easy to increase the disorder by small amounts at a
time by replacing some of the quadrupolar N$_2$ molecules by
spherical Ar atoms \cite{Scott,Manzhelii}. This allows one to
explore the interplay of disorder and frustration along with
interactions; these studies will be reported elsewhere.

It is generally believed that the orientational ordering in N$_2$
is nucleated at a temperature above $T_{\alpha\beta}\sim$36 K
although no experiment has indicated the temperature at which this
ordering begins \cite{Manzhelii}. The high temperature $hcp$ or
$\beta$-phase does not support long range orientational order but
there can still be some short range local ordering. In this
phase, N$_2$ is known to show hindered rotation due to the
incompatibility between the lattice and the molecular sizes
resulting in a complex free energy landscape. Recently we showed
that the dielectric constant ($\varepsilon(T)$) of pure solid
N$_2$ exhibited unexpected hysteresis in $\varepsilon(T)$ in the
audio frequency range whenever the sample is heated above an onset
temperature $T_h$ $\sim$ 42 K, in the presence of the bias
electric field \cite{PillaPL,PillaPhysica}. However, the
significance of the small external ac field was not obvious. In
this paper we present systematic dielectric as well as high
sensitivity heat capacity data for samples that are cooled either
in the presence or absence of an external ac electric field.
These results show that when field-cooled in a small ac field in
the audio frequency range, solid N$_2$ shows remarkable
glass-like memory and aging effects which are readily observable
within the laboratory time scales. In contrast, the zero-field
cooled samples do not show these effects.

The dielectric measurements were carried out using a three
terminal ac capacitance bridge having a sensitivity of two parts
per billion \cite{PillaRSI}. The high sensitivity heat capacity
measurements were carried out using an advanced dual-slope method
\cite{PillaHC}. Figure \ref{fildef5kV} shows the relative change
in $\varepsilon(T)$ in 4.2-55 K range (melting temperature is
$\sim$63 K). Curve (1), taken from ref.\ \ref{PL}, shows the
zero-field cooled result. The (lower) family of curves (2-4) show
$\varepsilon(T)$ when the sample is cooled in the presence of a
bias field of 5 kV/m at 1kHz, after warming up to $T_h < T_{max} <
48$ K. The difference in these curves is mainly due to the final
temperature ($T_{max}$) in the above range from which the sample
is cooled. As reported previously \cite{PillaPL,PillaPhysica},
once cooled down along any one of these curves, thermal cycles are
completely reversible along the same curve as long as the highest
temperature remains below $T_h$. The onset of strong hysteresis
upon warming above $T_h$ was noted in ref. \ref{PL}, but the
significance of the field-cooling was not appreciated. In the
(upper) family of curves (5-9), we show the results when the
sample temperature is raised to $\sim 53$ K, annealed at $50 < T
< 53$ K (for 10 to 12 h.), and then field-cooled in the presence
of the above bias field. The $\varepsilon(T)$ increases sharply,
reaching a maximum at $\sim 51$ K, and then decreases slowly. When
cooled from this temperature, the warm-up and cool-down curves are
no longer similar even in the $\alpha$-phase and are very
different from the zero-field cooled (curve (1)) or other lower
family of reversible curves (2-4). For $4.2 < T < 30$ K, the
$\varepsilon(T)$ decreases linearly with $T$ during warm-up
cycle, while it increases linearly with $T$ during cool-down
cycle. Also, we have observed spontaneous change in $\varepsilon$
at various temperatures indicated by $\mathbf{\downarrow}$ in
Fig.\ \ref{fildef5kV}. We note that the spontaneous changes in
$\varepsilon(T)$ occur at apparently random temperatures and
these are present in both the $\alpha$ as well as the
$\beta$-phases but \it only \rm during the cool-down cycle. Thus
the field cooled sample becomes trapped in metastable
configurations and the free energy barriers vary from very large
for the lower family of curves to sufficiently small for the upper
family of curves to allow random tunneling to lower states.
Remarkably, the value of $T_{\alpha\beta}$ is the same for all
curves to within 0.1 K although the change in $\varepsilon(T)$ at
$T_{\alpha\beta}$ can be different. If the system is left
isolated with the external field turned off and annealed above
$T_h$ for several hours and then cooled, $\varepsilon(T)$
retraces the lowest curve in Fig.\ \ref{fildef5kV}, independent
of the sample history. This shows that the sample can be brought
back to its original state by appropriate annealing, and that the
memory and aging effects are induced by cooling in the small ac
field, and not due to lattice dislocations or other defects.

When the sample temperature is further raised to $\sim$57 K, we
observed a massive, spontaneous change in $\varepsilon(T)$
($\sim2.5\times10^{-2}$) at 56.5 K shown in Fig.\
\ref{fildeflamda}. This temperature is highly reproducible to
within 0.1 K for various samples, though small hysteresis ($\sim$
1 K) in temperature is present between warm-up and cool-down
cycles (not shown). No anomaly near this temperature has been
observed in previous experiments, either in dielectric
measurements at microwave frequencies \cite{Kempinski} or in heat
capacity measurements \cite{Manzhelii}. The fact that such a
significant transition has not been observed before prompted us
to suspect that the transition may occur only in the presence of
the audio frequency electric field. However, because the
dielectric measurements always involve an external ac field, and
the bias field itself affects the measurement, it is not possible
to carry out a true zero-field cooled dielectric measurements at
these temperatures. We therefore carried out very sensitive heat
capacity measurements at constant volume \cite{PillaHC} for both
field cooled and zero-field cooled samples. Note that, unlike the
previous heat capacity data obtained using adiabatic methods, our
high sensitivity nonadiabatic method revealed, in the absence of
any external field (curve (1) in Fig.\ \ref{N2HC}), a new critical
behavior at $\sim$51 K in the heat capacity. This critical
behavior is in striking agreement with our dielectric data (inset
of Fig.\ \ref{fildeflamda}). In particular, in the absence of the
external electric field, $C_v(T)$ retraces curve (1) in Fig.\
\ref{N2HC} upon thermal recycling and the general behavior of
$C_v(T)$ near 56.5 K agrees very well with the previously
reported values \cite{Manzhelii}. When a uniform 10 kV/m, 1 kHz
field is applied across the sample from 10 K, the difference in
$C_v(T)$ is negligible up to the continuous transition near 51 K,
but significant deviation from the $C_v(T)$ of zero-field cooled
samples can be observed for $T > 51$K (see curve (2), Fig.\
\ref{N2HC}). When the above sample exposed to the external field
from 10 K is cooled down from 60 K and its heat capacity is
obtained once again during the warm-up cycle (while leaving the
field on through out), the observed $C_v(T)$ (curve (3), Fig.\
\ref{N2HC}) is larger than that of zero-field cooled sample in
the $\alpha$ as well as $\beta$-phases well above the sensitivity
of our apparatus. It should be noted that the structural
transition as well as the new continuous transition at $\sim$51 K
remain the same independent of the field. The data clearly shows
that in the field cooled samples, there exits a sharp
time-dependent rise in $C_v(T)$ near 56.5 K (indicated by
$\mathbf{\uparrow}$) which finally gives rise to a peak at this
temperature. Different curves in Fig.\ \ref{N2HC} near 56.5 K (in
curves (2) and (3)) obtained with 3 h.\ intervals indicate the
slow thermal evolution of $C_v$ with time. (In contrast, our
dielectric data (Fig.\ \ref{fildeflamda}) shows rapid change in
$\varepsilon$ at this temperature.) The absence of a peak in
C$_v$ for the zero-field cooled sample (curve (1), Fig.\
\ref{N2HC}) near 56.5 K indicates that the observed anomaly at
56.5 K is due entirely to the external electric field. Comparison
of curves (1) and (3) clearly demonstrates the strong effect of
the external ac field on the thermodynamic behavior of solid
N$_2$.

We have carried out several dielectric measurements to determine
the dependence on excitation field strength and frequency for
which solid N$_2$ shows this remarkable field-induced
nonequilibrium behavior. Several samples were initially
zero-field cooled down to 4.2 K, carefully annealed at $41.5 < T
< 43$ K for 6 to 8 h., and then annealed once again at $50 < T <
53$ K for 10 to 12 h. at various external field strengths and
frequencies. After cooling the samples to 4.2 K, a standard 100
V/m bias field at 1 kHz was applied to obtain the warm-up
$\varepsilon(T)$ data in the $\alpha$-phase for excitation field
strengths $<$ 100 V/m and frequencies $>$ 20 kHz. For other
excitation field strengths and frequencies, the bias field and the
excitation field are the same. This particular sequence is
followed because the ac capacitance bridge has the required
sensitivity only for bias fields larger than 100 V/m, and in the
audio frequency range \cite{PillaRSI}. Also, as pointed out
previously \cite{PillaPL}, once the sample temperature is below
$T_h$, the bias field has no observable effect on the
nonequilibrium behavior (and hysteresis) of the samples. From
these experiments we observed strong field-induced nonequilibrium
behavior for external electric fields stronger than $\sim$20 V/m
(peak-to-peak). The $\varepsilon(T)$ curves of the samples field
cooled in fields stronger than 20 V/m are very similar to those
shown in  Fig.\ \ref{fildef5kV}. For dc 1 kV/m as well as ac 2
V/m field cooled samples, the $\varepsilon(T)$ is only slightly
different from the zero-field cooled case (not shown). In
particular we do not observe spontaneous and random jumps in
$\varepsilon(T)$ and linear temperature dependence at these
temperatures, which characterize the nonequilibrim nature of the
field cooled samples. To find out the upper limit of the
frequency of the electric field for which we observe the
nonequilibrim behavior, we field cooled the sample in 5.2 kV/m,
90 MHz external uniform field. We observed no spontaneous and
random jumps in $\varepsilon(T)$ for this sample, but the small
hysteresis is still present and the $\varepsilon(T)$ curve is
close to that of the zero-field cooled sample. This indicates
that the glass-like states can be excited only at low frequencies
extending up to perhaps hundreds of kHz and for field strengths
greater than $\sim$20 V/m. This could be one of the reasons for
the failure of the previous microwave measurements
\cite{Kempinski} to observe the remarkable effect of the external
electric field on the thermodynamics of solid N$_2$.

We would like to point out that while for $T < 56.5$ K the field
cooled  $\beta$-phase N$_2$ shows remarkable aging behavior, for
$56.5 < T < T_M$, where lattice defects may be dominant due to
the proximity to melting, $\varepsilon(T)$ is linear as well as
reversible with thermal recycling (see Fig.\ \ref{fildeflamda}).
Here we would like to emphasize that the hysteresis values shown
in the various plots are accurate to within 1\%, and for
zero-field cooled samples, we observed no hysteresis or shift in
the absolute value of $\varepsilon$ at 4.2 K even after annealing
the sample for long time (few hours) at 50 to 55 K (without any
external field). This shows that the lattice defects present near
the melting temperature can not be the driving mechanism for the
aging and memory effects. The external electric field indeed is
responsible for the memory effects observed in solid N$_2$ whose
origin is not understood.

The strong effect of cooling in the presence of a small ac field
in the audio frequency range at such high temperatures is
puzzling. The audio frequency field corresponds to a temperature
scale of a few $\mu$K, and the strength of the electric field
coupling to the molecular polarizability is negligible compared
to the dominant electric quadrupole-quadrupole (EQQ) interactions.
Clearly the \it{process} \rm of orientational ordering is
disrupted by the presence of the field, although the field itself
is not strong enough to make a difference if the system is
already ordered (uniform field has no effect on EQQ). The
presence of short range ordering in the $\beta$-phase along with
the strong effect of field-cooling therefore implies that the
formation or growth of these clusters is inhibited by the
presence of the field.

The geometrical frustration generated by the symmetry
incompatibility of local and extended degrees of freedom results
in a thermodynamically large number of accessible ground states.
This macroscopic ground state degeneracy presents a new paradigm
with which one views condensed matter systems that form
glass-like states. The characteristic glass dynamics and aging
effects result from the relaxation among a large number of nearly
degenerate ground states. In the present case where in the
absence of substitutional disorder the frustration is due
entirely to the symmetry properties of the interaction, the
application of an applied electric field, although small, can
perturb the energy landscape of the interacting molecules and
result in the field-cooled memory effects. Thermal cycling to
temperature where the relaxation rate becomes sufficiently rapid
is required to erase the memory effect. In the case of pure
N$_2$, typical energy spacings are of the order of $\sim$1$\mu$K
\cite{Curl,Raich} (the $ortho$-$para$ spacing for local ordered
clusters) and ac fields of the order of 10 kHz are therefore most
effective in creating transfers among the energy landscape
profile. As a result of these transitions, a cluster of
orientatinally ordered molecules will find the molecular
orientations slightly changed from the locked ordering directions
leading to a partial destruction of the ordering, provided the
clusters are small enough. As observed, the effect is absent for
rf as well as pure dc electric fields. Upon field-cooling from
$T_{max} > 51$ K where large ordered clusters are not present,
the ac field modifies the landscape of metastable states and can
generate spontaneous tunneling to lower energy states (e.g.,
curve (8) in Fig.\ \ref{fildef5kV}). However, when we begin the
field-cooling from $T_{max} < 51$ K where a broad distribution of
ordered cluster sizes already exists, the small changes in the
orientation of the molecules can only affect the small clusters
(giving rise to hystersesis only). Thus the system can be
prepared in a variety of trapped metastable states (with small
differences in the ground state energies but high barriers),
leading to a free energy landscape with experimentally tunable
(large or small) barriers, by cooling from different temperatures
in the presence of an external electric field. In conclusion,
solid N$_2$ provides a unique opportunity to quantitatively
address a variety of questions in the rapidly evolving area of
aging and nonequilibrium phenomena in glass-like materials. This
work is supported by a grant from the National Science Foundation
No. DMR-962356.

\begin{figure}
\begin{center}
\leavevmode
\includegraphics[width=.95\linewidth]{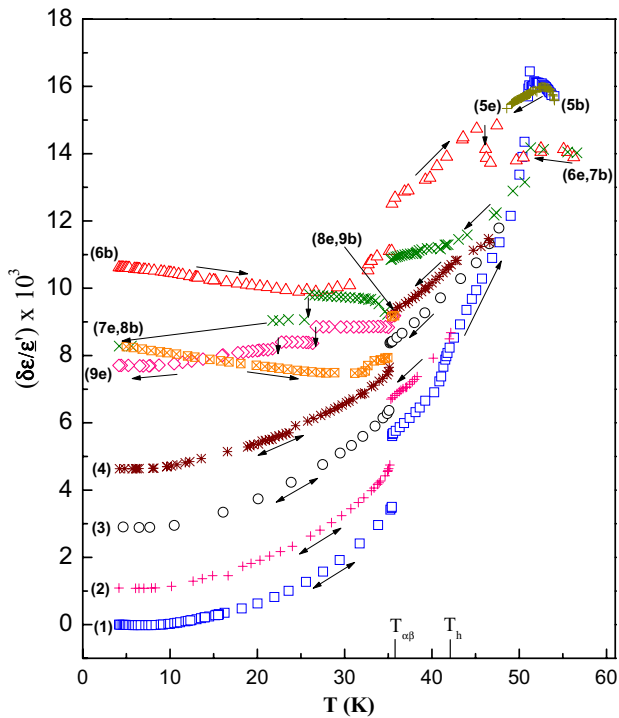}
\end{center}
\caption{$\varepsilon(T)$ of solid $^{14}$N$_2$ at 5 kV/m and 1
kHz excitation field relative to the value at 4.2 K on curve (1)
($\underline{\varepsilon'}$). $\underline{\varepsilon'} = 1.4255$.
Lower family of curves (1-4) show the gradual increase in
hysteresis with the highest temperature ($T_{max} <$ 51 K) reached
along each curve. Upper family of curves (5-9) are obtained after
annealing at $50 < T < 53$ K for 10 to 12 h.
$\mathbf{\rightarrow}$ along each curve indicates the direction of
the change in $T$ and the indices $b$ and $e$ indicate the begin
and end respectively of each curve. $\mathbf{\leftrightarrow}$
indicates the reversible change in $\varepsilon$ with thermal
recycling below $T_h$. $\mathbf{\downarrow}$ among the upper
family of curves indicates the spontaneous change in
$\varepsilon$ at random temperatures indicating the
nonequilibrium dynamics in the sample.}\label{fildef5kV}
\end{figure}
\begin{figure}
\begin{center}
\includegraphics[width=.75\linewidth]{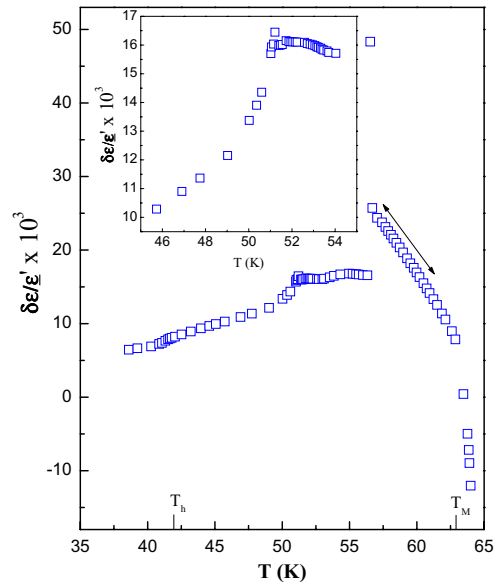}
\end{center}
\caption{$\varepsilon(T)$ of solid $^{14}$N$_2$ in the
$\beta$-phase. Above the sharp peak in $\varepsilon$ at 56.5 K,
$\varepsilon(T)$ is reversible with thermal recycling up to 61 K
indicating that lattice defects may not be the driving mechanism
for the nonequilibrim dynamics observed at lower temperatures.
Inset shows $\varepsilon(T)$ near 51 K in greater detail.}
\label{fildeflamda}
\end{figure}
\begin{figure}
\begin{center}
\leavevmode
\includegraphics[width=0.95\linewidth]{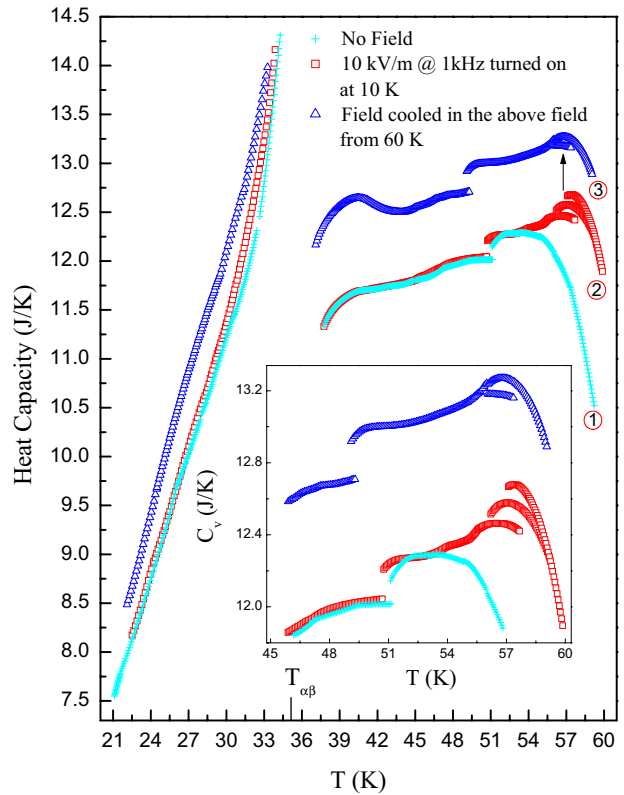}
\end{center}
\caption{Heat Capacity (C$_v$) of $\sim$ 0.36 mole solid
$^{14}$N$_2$. Different curves near 56.5 K (in curves (2) and (3))
obtained with 3 h.\ intervals indicate the slow thermal evolution
of $C_v$ with time.} \label{N2HC}
\end{figure}

\end{document}